\documentclass[journal,10pt]{IEEEtran}

\ifCLASSINFOpdf
\else
\fi
\usepackage{graphicx}
\usepackage{epstopdf}
\usepackage{stfloats}
\usepackage{bm}
\usepackage[cmex10]{amsmath}
\usepackage{algorithmic}
\usepackage{algorithm}
\usepackage{array}

\usepackage{mdwmath}
\usepackage{mdwtab}
\usepackage{multirow}

\usepackage{amsfonts}

\usepackage{amssymb}
\usepackage{balance}

\usepackage{color}

\begin{document}
%
\title{Towards Multi-Functional 6G Wireless Networks: Integrating Sensing, Communication and Security}

\author{Zhongxiang Wei,~\IEEEmembership{Member,~IEEE,}
        Fan Liu,~\IEEEmembership{Member,~IEEE,}
        Christos Masouros,~\IEEEmembership{Senior Member,~IEEE,}
        \\ Nanchi Su,~\IEEEmembership{Student Member,~IEEE,}
        and Athina P. Petropulu,~\IEEEmembership{Fellow,~IEEE}

\thanks{Zhongxiang Wei is with the School of Electronic and Information Engineering, at Tongji University, Shanghai, China. Email: z\_wei@tongji.edu.cn}
\thanks{Fan Liu is with the Department of Electronic and Electrical Engineering, at  Southern University of Science and Technology, Shenzhen, China. Email: liuf6@sustech.edu.cn}
\thanks{Nanchi Su and Christos Masouros are with the Department of Electronic and Electrical Engineering, at the University College London, London, UK. Email: \{nanchi.su.18, c.masouros\}@ucl.ac.uk}
\thanks{
Athina P. Petropulu is with the Department of Electrical and Computer Engineering, Rutgers University, NJ, USA. Email: athinap@rutgers.edu}


}

\maketitle

\begin{abstract}

Integrated sensing and communication (ISAC) has recently emerged as a candidate 6G technology, aiming to unify the two key operations of the future network in spectrum/energy/cost efficient way. 
ISAC involves communicating information to receivers and simultaneously sensing targets, while both operations use the same waveforms, the same transmitter and ultimately the same network infrastructure. Nevertheless, the inclusion of information signalling into the probing waveform for target sensing raises unique and difficult challenges from the perspective of information security. 
At the same time, the sensing capability incorporated in the ISAC transmission offers unique opportunities to design secure ISAC techniques. This overview paper discusses these unique challenges and opportunities for the next generation of ISAC networks. We first briefly discuss the fundamentals of waveform design for sensing and communication. Then, we detail the unique challenges and contradictory objectives involved in securing ISAC transmission, along with state-of-the-art approaches to address 
them. We then identify the opportunity of using the sensing capability to obtain knowledge of the targets, as an enabling approach against known weaknesses of PHY security. Finally, we illustrate a low-cost secure ISAC architecture, followed by a series of open research topics. This family of sensing-aided secure ISAC techniques brings a new insight on providing information security, with an eye on robust and hardware-constrained designs tailored for low-cost ISAC devices.

\end{abstract}

\IEEEpeerreviewmaketitle

\section{Introduction}


The 6G network, not only an improvement or extension of existing communication technology but also a great paradigm revolution, is envisioned as the new engine of the future intelligent world. It aims to build the foundation of the information infrastructures for the next decade. 
As a step ahead from connecting all the communication nodes in 5G networks, 6G will support ubiquitous communication, sensing, connectivity, and intelligence in a seamless manner \cite{Saad2021A}.
Among these exciting speculations on 6G networks, a consensus is that sensing, rising from an auxiliary functionality, will be a basic service of 6G networks, providing an extra dimension of capability of the network \cite{Liu2020Joint}.
This prompts the recent research interest of integrated sensing and communication (ISAC),  with significantly enhanced performance gains.
ISAC involves a single transmission that conveys information to communication users, while at the same time detecting targets. As such, it enables the integration of sensing and communication functionalities with a single transmission, a single device, and ultimately a single network infrastructure.
On one hand, by exploiting a common spectral, hardware platform and signal processing framework, ISAC is expected to significantly improve the spectral and energy efficiencies, addressing spectrum congestion while reducing hardware and signaling costs.
On the other hand, ISAC exploits the co-design between the two functionalities, enabling communication-aided
sensing and sensing-aided communication. Hence, it considerably improves sensing and communication performance, especially in localization-, environment-aware scenarios.
Benefiting from the above merits, ISAC is able to build the potential of many emerging applications, including and not limited to intelligent factory, smart city/home, connected cars, and e-Health systems, as illustrated in Fig. 1.


\begin{figure*}
	\centering
	\includegraphics[width=3.9in]{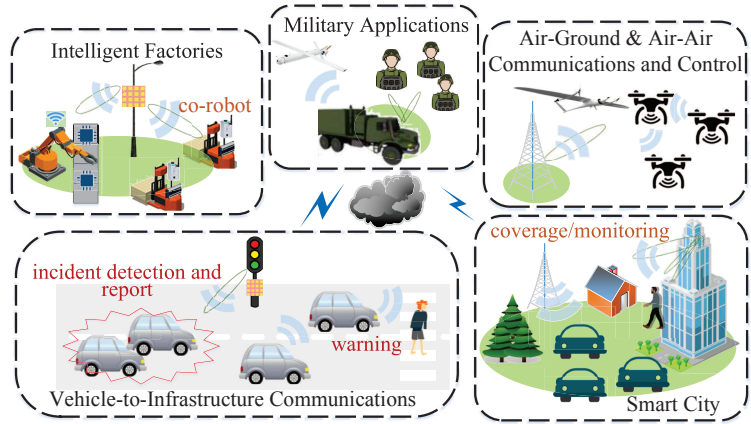}
    \caption{
     Sensing-aided communication and communication-aided sensing  techniques are preferable through the upcoming paradigm of ISAC, which greatly prompts the applications of intelligent factories, smart cities/homes, vehicle-to-infrastructure communications, military communications, etc.}
    \label{fig:anonymous_mag} 
\end{figure*}

Nevertheless, due to the spectrum sharing mechanism as well as the broadcasting nature of wireless transmission, ISAC raises unique security challenges. The inclusion of information messages into the radar probing signal makes the communication susceptible to eavesdropping from the  targets.
Indeed, the target that is being sensed can potentially exploit the information-bearing signal, and detect the confidential message intended for the information users \cite{Su2021Secure}. This raises a unique and very interesting tradeoff where, on one hand the transmitter wishes to illuminate the target by focusing  power towards its direction, and on the other hand, it has to limit the useful signal power at the target to prevent eavesdropping.

A possible solution is to apply cryptographic techniques at high layers of the network stack to encrypt the confidential data prior to transmission.  
However, these upper layer secure solutions have several limitations, including and not limited to the tedious secret key distribution/management/maintenance processes, unprovable security performance in the presence of a computationally strong eavesdropper (Eve), and difficulty in identifying a compromised secret key.
Importantly, even if the information message is encrypted with higher layer techniques, the presence of a communication link can still be detected by the eavesdropping target, which could jeopardize the transmitter's key parameters such as its ID, location, and other sensitive information. Accordingly, physical layer (PHY) security, an information-theory-based methodology that resides at lower layer, 
appears as a complementary approach for securing wireless transmission. It has been shown that, by exploiting the channel disparity between Eves and legitimate users (LU)s, it prevents the signal detectability at  Eves directly; 
it does not rely on the higher layer cryptographic techniques as well as the limitations of Eves' computational resources;
it releases the difficulties in the secrecy keys distribution and management, and thus is able to provide security even though the secret key may have been compromised \cite{Bloch2021An}.
Among the existing PHY techniques, some use multiple transmit antennas for secure beamforming and artificial noise \cite{Wei2020Energy} \cite{Dong2009Improving}. Some require help from relay,  LUs or third party for cooperative jamming. Some exploit the PHY attributes of communication links, such as channel impulse response, received signal strength indicator, carrier frequency offset, in-phase quadrature-phase imbalance, for authenticating LUs.  Some address PHY security at the cost of reduced communication efficiency, such as constellation rotation and noise aggregation, as summarized in Table I.



Despite the decades long research on PHY security, the major limitation of a large class of PHY security solutions stems from the need of knowing the  Eves' channels, or direction as a minimum.
Interestingly, the sensing functionality of ISAC brings a new opportunity for secure design.  
That is, by proactively sensing an ambient Eve and designing a dedicated waveform to suppress the receive quality-of-service (QoS) of the Eve, the additional sensing functionality can serve as a strong support to facilitate the provision of security.
Motivated by the aforementioned issue, the purpose of this article is to overview the sensing-aided secure design in accordance with the characteristics of ISAC.  
Starting from briefly introducing the fundamentals of ISAC systems, we first examine a novel secure ISAC design, where its waveform is judiciously designed with the aid of the integrated sensing functionality. Then, we further discuss a practical robust secure ISAC design, where the knowledge of the target and communication users is imperfectly obtained.
Further, a hardware-efficient secure ISAC  architecture, based on the concept of directional modulation is reviewed.
Open challenges are then identified, before concluding this article.

\begin{table*}
\centering
	   \caption{A Brief Summary of The Existing ISAC and PHY Security Design.}
	\includegraphics[width=6.4 in]{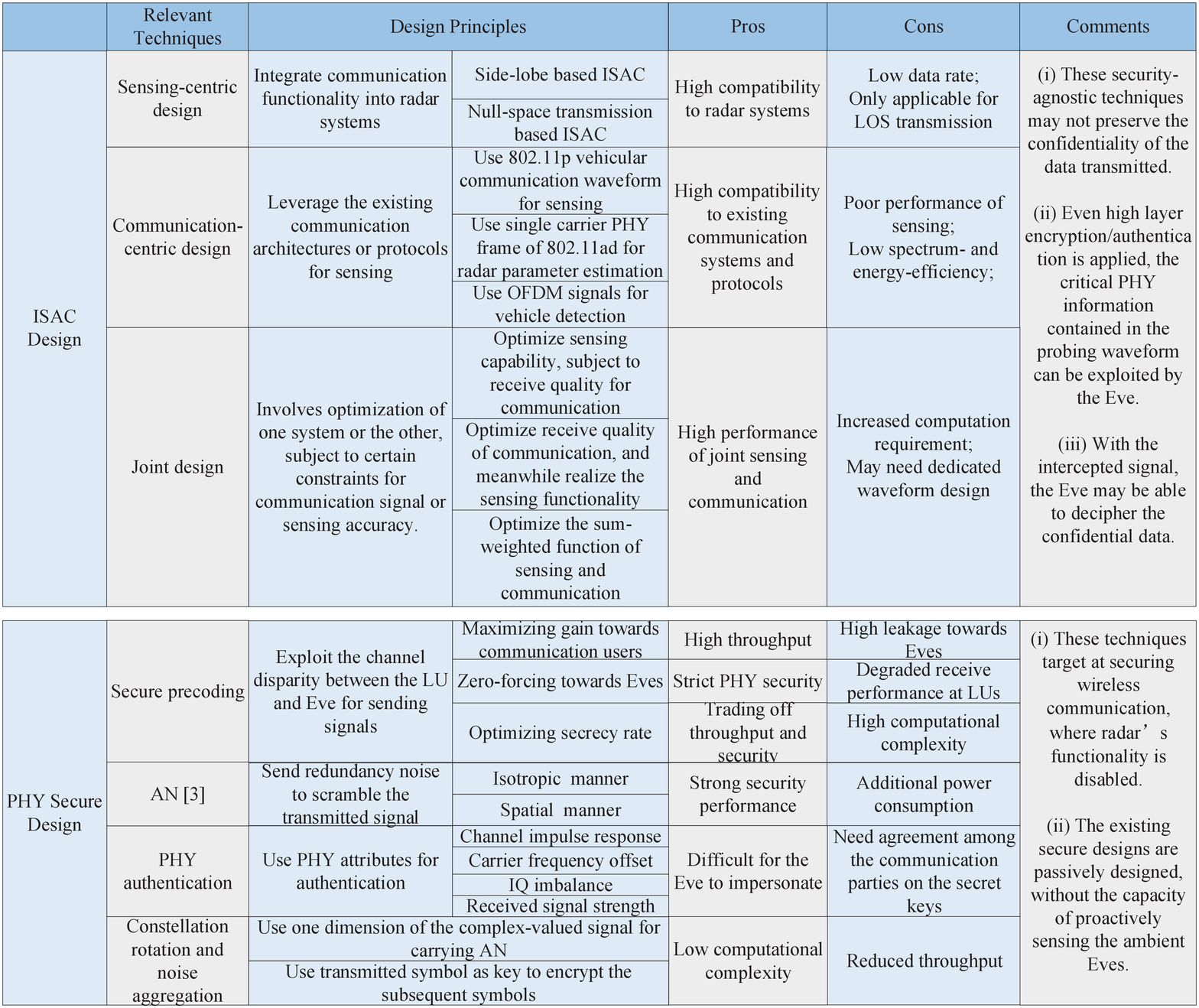}
    \label{fig:summary} 
\end{table*}

\section{The Fundamentals of ISAC} 

As an early stage of  ISAC, 
communication and radar spectrum sharing (CRSS)  has been investigated from the perspective of spectrum sensing, dynamic spectrum access, and mutual interference mitigation \cite{Li2016Optimum}. The generic aim of CRSS is to develop efficient spectrum utilization and interference management mechanisms, so that the two systems can share the spectrum without significantly interfering with each other. Hence, CRSS needs to share design parameters between two systems, or relies on the coordination from a control center.
As a further step, ISAC is able to realize not only the spectral coexistence, but also the shared use of hardware flatform and even network architecture, as shown in Fig. 2. 
Hence, radar and communication functionalities are simultaneously realized with efficient spectrum utilization as well as low hardware cost. Most importantly, this regime enables a deeper integration of the two functions, where sensing and communications are not perceived as separate end-goals, instead lending themselves to communication-aided sensing and sensing-aided communication. 
Let us start by demonstrating the fundamentals of ISAC, and then elaborate on the secure ISAC transmission.



\subsection{Sensing Basics}

While communication can be characterized  as to accurately and efficiently recovering the information delivered by the transmitter at the receiver, sensing refers to extracting information from the radio waves reflected/scattered by the target(s) of interest, which can be realized either in an active or passive manner. Consequently, the useful information is contained in the target return, rather than in the sensing waveform itself, which is distinctly different from wireless communications. 

In general, sensing tasks can be split into three different categories, namely, detection, estimation, and recognition. Detection aims to make binary decision on the status of a target, typically on its presence or absence. Estimation operation, on the other hand, infers useful parameters about the target from noisy returns, e.g., range, velocity, and angle. Finally, recognition refers to identifying what the sensed target is. 
Profiting from the abilities of detection, estimation and recognition, ISAC is able to support more exciting applications in 6G networks, such as environment mapping and semantic positioning.
It is noteworthy that these tasks are built upon different theories.
Detection and estimation rely on detection and estimation theory, which is more relevant to PHY layer signal processing. Recognition is based on imaging and learning theories, which is at the application layer \cite{Liu2020Joint}. In this article, we focus on PHY signal processing of sensing which is  
exploited for assisting secure waveform design, and set aside higher layer designs as a future work.

Waveform design plays a significant role for enhancing both the communication and sensing performance. Interestingly, since that sensing and communication performances are evaluated by different key performance indicators, ISAC waveform design should take different metrics into consideration for implementing the dual functionalities. This typically incurs conflicting design objective between sensing and communications, which needs to be carefully balanced as detailed in the next subsection. 

\subsection{Waveform Design for ISAC}

ISAC waveform designs can be categorized into sensing-centric design, communication-centric design, and joint  design, as examined in the following.

\begin{figure*}
	\centering
	\includegraphics[width=4.1 in]{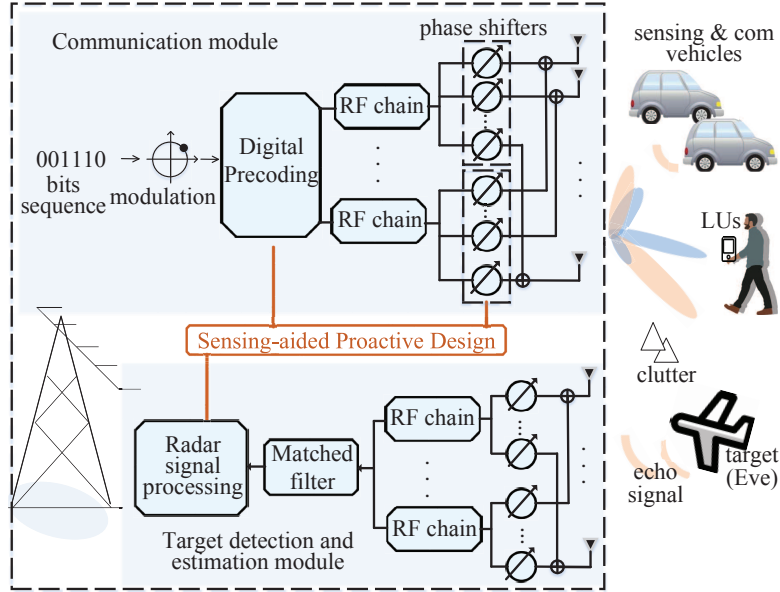}
    \caption{A generic ISAC architecture based on the philosophy of joint design, where its waveform is dedicatedly designed to approach an ideal sensing beampattern, and at same time carry confidential data with a high receiver-end SINR  for communication. On the other hand, while the reflected echo waveform is analyzed for target detection and estimation, the sensing results also  assist secure waveform design in a proactive manner.}
    \label{fig:anonymous_mag} 
\end{figure*}

\textit{Sensing-Centric Design}:
The philosophy of the  sensing-centric design is integrating  communication messages into a classical sensing waveform, and hence it has a high compatibility to the existing radar architecture.
The early stage of sensing-centric design involves pulse interval modulation, where the interval between radar pulses is utilized for communication. 
Or, the sensing-centric design can leverage the concepts of index modulation \cite{Huang2020MAJoRCom} and generalized spatial modulation for waveform design \cite{Xu2021A}.
In contrast, another sensing-centric design philosophy is to  sense the target in the mainlobe of the radar beampattern, while transmitting useful information  in the sidelobes. 
One can simply modulate its sidelobe using amplitude shift keying (ASK) or phase shift keying (PSK), where the communication symbols are represented by the different levels of power, or the different phases of the received signals \cite{Hassanien2016Dual}.
Nevertheless, since a communication symbol is generally
embedded into either a single or several radar pulses, the sensing-centric design results in a low data rate that depends on the pulse repetition frequency (PRF) of the radar, which is well below 5G/6G requirements.
Also, the sidelobe based schemes only enable line-of-sight (LoS) transmission.  This is because for a multi-path communication channel,
the communication quality will be  distorted by the dispersed
signals arriving from Non-LoS (NLoS) paths, where all the
sidelobe and the mainlobe power may contribute equally.



\textit{Communication-Centric Design}:
The philosophy of the communication-centric ISAC is to leverage the standardized  communication waveforms, protocols and architectures for sensing.
For example, pilot signals and frame preambles that have high auto-correlation properties and are typically used for channel estimation or multi-user access, have been recently employed for sensing targets.
Also, standards-relevant communication waveforms have been used  for sensing targets and obstacles in vehicular applications.
The IEEE 802.11p vehicular communication waveform was  used
for radar sensing at the 5.9 GHz band, but its detection range and
velocity estimate fail to meet the desired accuracy requirements of automotive radar applications. 
An alternative approach  is to leverage the IEEE 802.11ad protocols, where the preamble of a single carrier PHY frame is used for
 parameter estimation in vehicular applications \cite{Kumari2021Adaptive}.
Overall, these communication-centric ISAC designs can realize sensing functionality without sacrificing the communication performance, thereby leading to high data rate. 
However,  the pilot signal, frame preambles and communication waveforms are not dedicatedly designed for sensing, and  the sensing operation is constrained inside the communication resources.  
Accordingly, the main drawback of the communication-centric designs lies in the poor,  scenario-dependent and difficult-to-tune sensing performance. 

\textit{Joint Design}:
More recently, joint design of radar and communications has attracted much attention, because it enables high performance of both sensing and communication. 
In joint-design ISAC approaches, the beampattern is dedicatedly manipulated to achieve an ideal radar beampattern, while ensuring a high receiver-end signal-to-interference-plus-noise ratio (SINR)  for communication \cite{Fan2019Toward}. 
In general,  joint design involves  joint optimization of both functionalities and enables  scalable performance trade-offs between them.
For example, the  waveform can be designed to optimize the sensing performance of the radar, subject to satisfying certain reception quality constraints for communication. 
Alternatively, one is able to optimize the reception quality of LUs and meanwhile ensure the radar's functionality. 
To strike a trade-off between sensing and communication performance, the sum-weighted sensing and communication quality can also be exploited as an objective function, further leading to a Pareto-optimality of the multi-objective optimization \cite{Liu2020Joint}. 
Evidently, the joint design enables flexible use of time, frequency, and spatial resources, thereby achieving both high throughput as well as sensing reliability. 
Most importantly, the joint design prompts the ultimate integration of sensing and communication, with a common  protocol and  network infrastructure.

\section {From Dual-Functional to Multi-Functional: Integrating Security into ISAC}

In this section, we explore the synergies between ISAC  and security, where the  sensing functionality is judiciously utilized for benefiting the provision of information security.

\subsection{The Unique Security Challenges and Opportunities of ISAC}

As mentioned, the ISAC transmitter needs to focus its transmit power towards  directions of interest to obtain a good estimation of
the targets. It implicates that the radar target has a high reception SINR on the embedded confidential signal, which significantly increases  the susceptibility of information to eavesdropping by the  targets.
As a result, the  target may be able to intercept the information-bearing signal, and  extract the confidential message intended for the communication users. 
This raises a unique security challenge for ISAC systems, requiring to carefully strike a trade-off between concentrating  direct power towards the target's direction for sensing and limiting the useful signal power at the target to prevent eavesdropping.
To facilitate high quality of target sensing while keeping the confidential signal unbreakable by the target, in the following we examine the recent research line of sensing aided secure ISAC techniques.

Since the ISAC transmitter is able to sense the angle of the target and estimate its roundtrip channel, this can be  treated as the wiretap channel from the perspective of information security.
Hence, with the proactively obtained angle and the wiretap channel, the ISAC transmitter is then able to calculate the  eavesdropping SINR $\bm{\Gamma}_{\mathrm{E}}$ of the target  prior to transmission. 
This information for the potential Eve can be used to design a number of secure transmission approaches such as secure beamforming, artificial noise approaches and jamming, cooperative security, amongst many others. The unique ISAC transmission however requires these to be redesigned for the dual functionality. As an example, in designing a secure dual functional transmission, the unique aim of secure ISAC transmission can be maximizing the  echo signal's signal-to-clutter-plus-noise ratio (SCNR) $\bm{\Psi}$ at the ISAC's receiver for sensing performance, while limiting the eavesdropping SINR $\bm{\Gamma}_{\mathrm{E}}$ at the target and at the same time guaranteeing the intended signal's SINR $\bm{\Gamma}_{\mathrm{LU}}$ at LUs above a certain threshold. This equivalently enhances the value of the secrecy rate, calculated as the achievable rate difference between the LUs and target. 
Alternatively, one can maximize the secrecy rate and meanwhile ensure  the echo's  SCNR  performance at the ISAC receiver for guaranteeing the radar's functionality.
The critical aspect, unique to ISAC, is that now the angle of the sensing beam involved in the SCNR objective is the same as the angle involved in the eavesdropping constraint. These contradicting constraints make the secure ISAC design difficult and raise interesting trade-offs between the sensing performance and the security. On one hand, the ISAC transmitter aims to maximize the SCNR to illuminate the target for detection, while on the other hand it desires to minimize the eavesdropping SINR at the same direction, as shown in Fig. 3(a).



Designing of  a sensing-aided secure waveform is not convex in nature, due to the fractional-structured SINR constraints of the sensing and communication functionalities, as well as the rank-1 constraint of the beamforming matrix. Generally, the fractional SINR constraints can be transformed into subtract-structured optimization problem \cite{Su2021Secure}. Then, dropping the  rank-1 constraint, the globally optimal solution can be obtained by iteratively solving a sequence of semi-definite programming. 
Note that in the rare case that the target and LUs are in the same direction and have strong LOS channels,  ensuring security at the PHY layer is extremely challenging and authentication and encryption secure techniques are still needed at the higher layers \cite{Bloch2021An}.

\subsection{Robust Secure ISAC Waveform Design}

In practice, a target's position may not be always perfectly obtained,  due to sensing error and finite detection resolution. 
Typically, the sensing error is introduced by noise while the finite detection resolution is due to limited array aperture and bandwidth.
For example, given $N$ antennas fabricated in uniform linear structure with half-wavelength spacing, the angular resolution is approximately calculated as $\frac{2}{N}$ (in rad) \cite{Liu2020Joint}, which means the targets within that angular interval can not be detected individually.
When the target’s position can only be roughly sensed within an angular region, a wider beam needs to be formulated towards that region to avoid missing the target. 
However, focusing the beam to a region of space inevitably leads to an increased possibility of the information leakage, giving rise to a need for robust secure waveform design.

\begin{figure*}
	\centering
	\includegraphics[width=6.8 in]{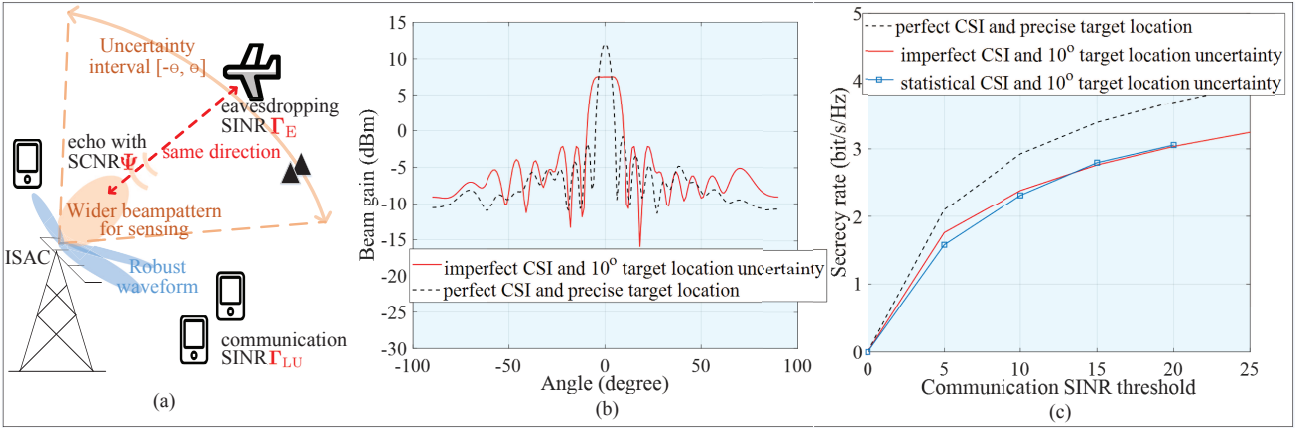}
    \caption{
    (a) A  practical scenario, where the target’s position is roughly sensed within an uncertainty interval, and the LUs' channels are also imperfectly known.
    (b) The width of the beampattern is adaptively manipulated in different scenarios for sensing the target. 
    (c) By proactively sensing the target, a high level of secrecy rate is achieved. }
    \label{fig:simulations} 
\end{figure*}

When the target is known to be located with certain angular region of space, obtaining a robust secure waveform can be obtained by minimizing the sum of the target's reception SINR at the possible locations in this angular interval. This formulation  can be handled in a similar way as in the ideal case. 
More importantly, the achievable rate of the target is upper-bounded, thus guaranteeing  information security. 
On the other hand, when the LUs' channels are not perfectly known by the ISAC transmitter,  the channel estimation error can be generally formulated using bounded or un-bounded error models \cite{Wei2021Multi}. 
To be specific, the former model assumes that the norm of  error is bounded while the   statistical distribution of the error is unknown. In contrast, the latter model assumes that the estimation error follows Gaussian distribution, and hence its norm is not bounded.
With the bounded or un-bound error model, computing the LUs' SINR  constraints can be further transformed into deterministic or probabilistic  robust optimization problem, which can be readily handled by a series of established optimization tools \cite{Wei2021Multi}.

Let us consider the application scenario shown in Fig. 3(a), where the angular interval of target location uncertainty is  $[-5^{\circ},5^{\circ}]$, while the channel estimation error of the LUs follows Gaussian distribution with variance 0.05.
There are 4 LUs and their required SINR threshold is 40 dB. The  power budget is 20 dBm. The objective of the secure waveform optimization is to suppress the targets' reception SINR, subject to per-LU's SINR requirement, power constraint, 
while ensuring the resulting waveform approximates  the desired sensing beampattern.
As observed in Fig. \ref{fig:simulations}(b), a narrow beampattern is obtained when the target location is precisely sensed at the ISAC transmitter. 
Leveraging the proactively obtained location, the transmitter is able to manipulate the dissipated waveform to suppress the eavesdropping SINR of the target, thereby improving
the secrecy rate  in Fig. \ref{fig:simulations}(c). 
When the target's location can only be imperfectly sensed, a wider beampattern is formed, directing the same power over the region of possible target location; in this case the power gain of mainbeam reduces.
Nevertheless, by suppressing the sum of the target’s SINR at the possible locations in the  angular interval, a high level of secrecy rate is achieved, even if the ISAC transmitter only knows the statistics of the LUs' channels.

\subsection{Hardware Efficient Secure ISAC  Design}

In real scenarios,  hardware limitations may jeopardize the sensing and communication performance, and importantly the security of the transmission. A recent abundance of hardware efficient techniques that have been developed for communication-only systems can be leveraged to design hardware-informed secure ISAC transmission \cite{Wei2021Secure}.
In this section,  the validity and extensions of secure ISAC into hardware-constrained scenarios are investigated.

The implementation of fully-digital MIMO ISAC,  requiring a radio frequency (RF) chains per antenna element, is prohibitive from both  cost and power consumption perspectives. 
On the feasibility of secure waveform with high hardware efficiency, one approach is to reduce the RF chains through analog architectures that involve phase shifters (PS)s and/or switchers, termed as hybrid ISAC systems, as illustrated in Fig. 2.
This approach involves  low-dimensional baseband beamforming, followed by high-dimensional analog beamforming.
However, in both  fully-digital or hybrid ISAC, the required number of RF chains is no smaller than the total number of data streams for multi-user access.
To remove the expensive and power-consuming RF chains and the digital-to-analogue converters (DAC)s, a more hardware-efficient secure ISAC technique, built on the concept of directional modulation (DM) is emerging, where  PSs or parasitic antennas are used as main components in the transmitter \cite{Wei2021Secure}.

Aided by the DM technique, symbol modulation happens at the antenna level instead of  the baseband level, and the received beam pattern at the LUs is treated as a spatial complex constellation point.  
In particular, the constructed signal of the LUs does not necessarily align with the intended symbols, but can be pushed away from the detection thresholds of the signal constellation, based on the concept of constructive interference (CI) regions \cite{Wei2021Multi}.  
The resultant increased distance with respect to (w.r.t) the detection
threshold directly benefits the LUs' reception quality.
An example is illustrated by Fig. 4 for quadrature
phase shift keying (QPSK) QPSK and 8PSK.
Since the decision thresholds for QPSK  are the real
and imaginary axes,  the constructed symbols (denoted by blue dots) at the LUs can be  judiciously  pushed  away from both the real and imaginary axes.
In a similar fashion, the symbols can be constructed for the LUs with 8PSK or higher-order modulations.
On the other hand, with the proactively obtained Eve's information, we can intentionally push the  Eve’s received symbols (denoted by red dots) into destructive regions of the signal constellation.
This destructive interference impedes the  Eve’s intercepting behavior at a symbol level, and hence the volume of the symbols being intercepted in any block is significantly reduced.

\begin{figure*}
	\centering
	\includegraphics[width=4.6 in]{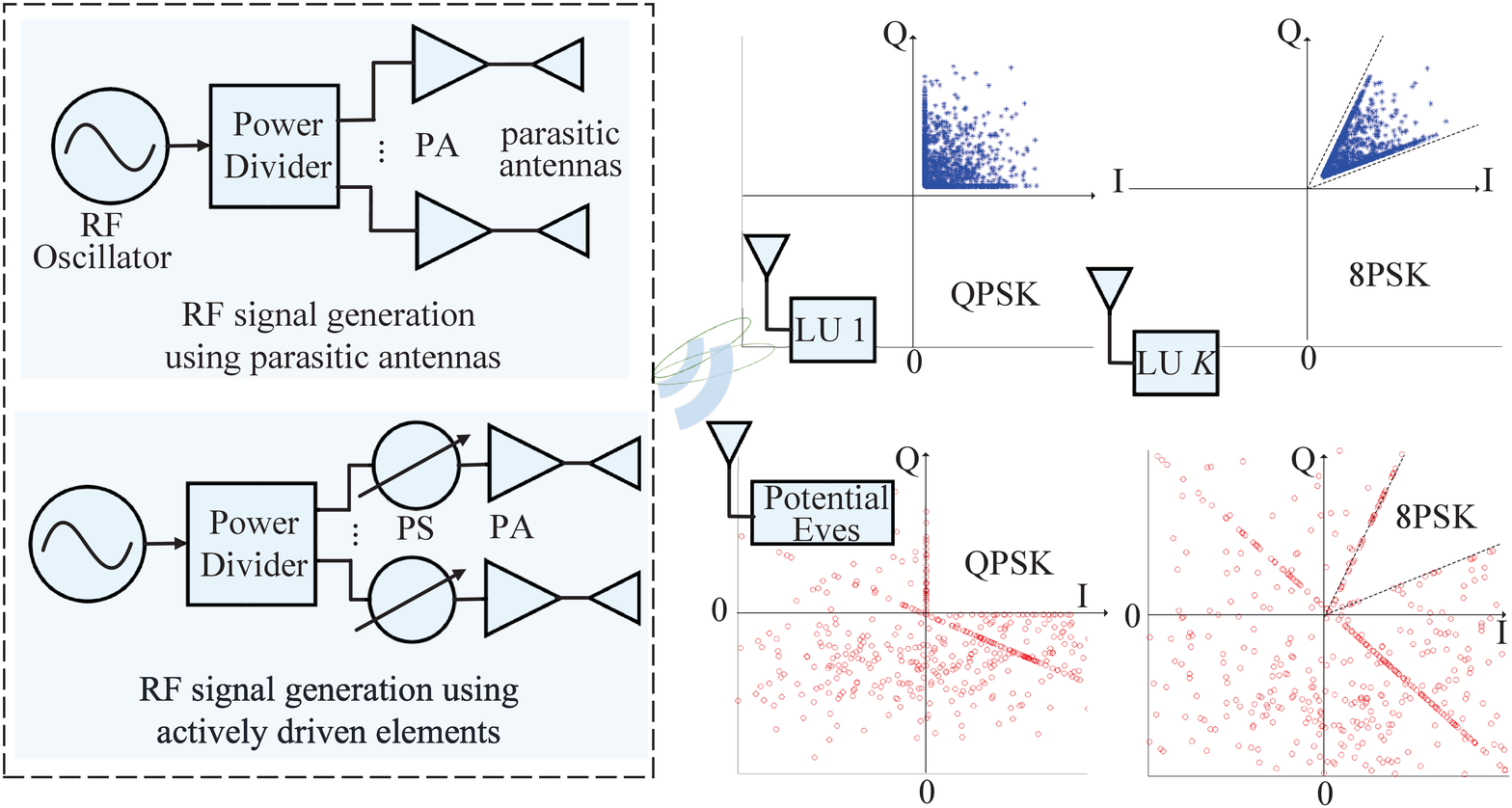}
    \caption{RF signal can be directly modulated at the antenna level by use of driven elements or parasitic antennas. The LUs' received symbols (denoted by blue dots) are shifted into CI regions directly. Assisted by the sensing functionality, the  Eves' received signal (denoted by red dots) can be intentionally located into destructive regions, resulting in a high symbol error rate.}
    \label{fig:anonymous_mag} 
\end{figure*}

\section{Open Challenges and Future Works}

ISAC-relevant design is still broadly open, and the remaining challenges can benefit from the communications literature.


\textbf{Radar Location and Identity Privacy-Preserving Design}: 
Though sharing both spectrum and hardware platform in ISAC has attracted much attention, the CRSS system still has its market on the evolution road of the radar and communication co-existence.
Designed to control mutual-interference, there are always
parameters passed to one system that contain implicit information
about the other. This raises privacy concerns for the
two systems, and especially for the military radar. 
With the knowledge of the waveform scheme, 
a legitimate but curious  communication transmitter is possible to  reverse-engineer the radar parameters, further localizing the radar. Recent research has unveiled some machine-learning based schemes, which exploit the information contained in the precoder to infer the radar’s location \cite{Dimasa2019On}. 
As a result, how to exchange design parameters between radar and communication units without loss of each other's privacy, while  maintaining a minimum level of mutual interference remains an open challenge.

\textbf{Secure ISAC Design for 5G/6G KPIs}: 
The existing ISAC designs generally consider  SINR or end-to-end throughput as metrics for evaluating communication performance. 
In recent years, ultra reliable  low latency (URLLC), and machine-type communications have received a lot of attention in practical applications.
Together with the requirement of   information security, those  applications also involve new metrics and specific transmission  protocols, such as latency, reliability, massive access, short packets, and so on.
Rethinking  secure ISAC techniques to align with these stringent requirements, and also to maintain a low level of complexity and overhead is an open and fertile area of research.

\textbf{On  Compatibility  of Secure ISAC  and 5G NR}:
5G new radio (NR) has standardized a series of waveforms, including but  not limited to CP-OFDM, filter-OFDM, UF-OFDM, DFTS-OFDM,  and FBMC-QAM. 
Also, 5G NR has also proposed adaptive wireless interface configuration, such as changeable frame structure and adaptive 15 KHz-120 KHz carrier spacing.
Hence, the compatibility of the secure ISAC with  5G NR remains an open challenge. 
With different communication environments and specific performance requirement, how to leverage the flexible waveform specification and wireless interface configuration? Essential work is needed to bridge the gap between theory and implementations.

\textbf{Network Level ISAC Design and Secure Performance Analysis}: 
There has been extensive research on the networking design and performance analysis for generic communication systems, where SINR, converge probability, outage probability and ergodic system capacity are analyzed and optimized in a systematic manner, based on the  stochastic process and calculus.
This network level investigation  advises networking planing and  engineering design with an eye to the interests of the whole system.
While the existing ISAC-related research is investigated in simple scenarios,
considering the heterogeneity and high nodes density in future communication systems, the systematic ISAC design and  performance analysis   need fundamental  research.  


\section{CONCLUSIONS}
This article has discussed the exciting intersection of ISAC and security.
Starting from the fundamentals of the ISAC, we first have discussed the methodology of the waveform design for joint sensing and communication.  
Then, we have examined the sensing-aided secure ISAC techniques  to prevent the confidential signal embedded in the probing waveform from being eavesdropped upon by the sensing target. 
Finally, the recent interests in energy- and hardware-efficient ISAC has been reviewed, where the  RF chains are replaced by PSs or parasitic antennas for reducing power and hardware cost. 
This family of sensing-aided secure ISAC design offers a broad field of preserving information security in a proactive manner, which holds the promise of exciting research in the years to come.

\ifCLASSOPTIONcaptionsoff
  \newpage
\fi

\end{document}